# Synthesis of nZVI /PVP Nanoparticles for Bioremediation Applications

**Anatoli Sidorenko[1], Tatiana Gutul[1], Mine Gül Şeker[2], Tuğça Arit[2], E. Gutul[3], Anatoli Dimoglo[4], Ashok Vaseashta[5*]**

[1]Institute of Electronic Engineering and Nanotechnologies 'D.Ghitu' ,3/3Academie, MD-2028 Chisinau,  Republic of MOLDOVA; anatoli.sidorenko@kit.edu tatiana.g52@mail.ru

[2]Gebze Technical University, Department of Molecular Biology and Genetics, Çayirova, Kocaeli, 41400, TURKEY; gul@gtu.edu.tr tugcearit@gtu.edu.tr

[3]Nesmeyanov Institute of Organo-element Compounds, Russian Academy of Sciences (INEOS RAS), Vavilov str. 28, Moscow, 119991, RUSSIA; gevgenii@mail.ru

[4]Düzce University, Environmental Engineering Department, Konuralp, Düzce, 81620, TURKEY; anatolidimoglo@duzce.edu.tr

[5]International Clean Water Institute, Manassas, VA USA; Riga Technical University, Riga, LATVIA; Institute of Electronic Engineering and Nanotechnologies 'D. Ghitu', Chisinau MOLDOVA; prof.vaseashta@ieee.org

* Correspondence: prof.vaseashta@ieee.org

**Abstract:** The objective of this investigation is to study zero-valent iron (ZVI) nanoparticles (NP) for bioremediation applications. The ZVI-NPs were fabricated by chemical reduction using a ferrous salt solution with poly(N-vinylpyrrolidone) (PVP), used as a stabilizer. The synthesis was conducted with and without ultrasonic treatment. The ZVI nanoparticles prepared as described here, were characterized using scanning electron microscopy (SEM), transmission electron microscopy (TEM), X-ray powder diffraction (XRD) analysis and Fourier Transform Infrared Spectroscopy (FTIR). Experimental observations demonstrate that depending on synthesis conditions and co-ordinations of the stabilizer, nanoparticles with different morphologies are formed. Colloidal solutions of the synthesized nanoparticles were used in antimicrobial activity tests and biofilm formation assays for nine different control microorganisms: *Escherichia coli* (ATCC 25922) representing Gram negative bacteria, *Pseudomonas aeruginosa* (ATCC 15692), *Enterococcus fecalis* (ATCC 29122), *Proteus vulgaris* (laboratory isolates), *Staphylococcus aureus* (ATCC 29213) representing Gram positive bacteria, *Bacillus cereus* (DSMZ 4312), *Bacillus subtilis* (ATCC 6633) used as a soil microorganism, and *Candida albicans* (ATCC 10231) selected as a representative of fungal microorganisms. The formation of a biofilm and the absence of an antimicrobial effect are observed for PVP-stabilized ZVI nanoparticles that were synthesized without ultrasonic treatment. The results from this investigation demonstrate that ZVI nanoparticles have bioremediation potential and can be effective, efficient and sustainable for treatment of contaminated surfaces. Hence, a selected bioremediation application of nZVI-PVP are discussed.

# 1. Introduction

The environmental pollution due to anthropogenic contaminants is increasing and remains as one to the top global challenges of the 21$^{st}$ century to be addressed by scientists and policymakers. Anthropogenic changes and their impact on the natural environment are often not well understood due to its evolution cycle, and challenges and complexity related to differentiating natural variation from anthropogenic changes. With recent advances in nanomaterials, the pace of disruptive technological innovation has advanced from linear to parabolic. The material aspect of several innovations is based on novel and unique characteristics of nanomaterials due to its reduced dimensions and large surface area per unit mass (Vaseashta, 2005) which can be useful in hazardous waste site remediation and contaminant reduction. Nanoscale zero-valent iron (NZVI) particles can rapidly transform many environmental contaminants to benign products and are a promising *in-situ* remediation agent. Due to toxic impacts associated with pesticides, many recent studies have focused on the decomposition of pesticides in aqueous media and soils using nanoparticles (NPs), since they can be injected where remediation and reclamation of soils and lands is of utmost necessity.

A strategic approach to remediation and reclamation of soil investigation is based on the optimal use of natural processes that involve self-purification and self-regeneration, such as bioremediation and other similar techniques using new technologies, such as iron-containing NPs (magnetite ($Fe_3O_4$), goethite $FeO(OH)$, hematite $\alpha$-$Fe_2O_3$, and zero-valent iron (ZVI) $Fe^0$) exhibiting pronounced redox properties (Asadishad et al. 2017; Basnet et al. 2016; Yin et al. 2019). It has been proposed that these NPs should be used for the remediation of both organochlorine pesticides and nitroaromatic pesticides. There are only a few studies on nitro pesticides that include trifluralin and are mostly focused on the degradation of the pesticides in water (Xie et al. 2017).

Here, a special experimental niche is applied using ZVI NPs, which demonstrated an extremely high efficiency in treatment to remove heavy metals and persistent organic pollutants (POPs) in wastewater. In particular, the use of iron NPs and nanocomposites is based on its remediation of residual pesticides (organochlorine dichlorodiphenyltrichloroethane (DDT), dichloro- diphenyldichloroethylene (DDE), dichlorodiphenyldichloroethane (DDD), and hexachloro-cyclohexane (HCH) having potential toxic effects on soil microorganisms (Rostek et al. 2018).

Several researchers have concluded that the microbial community in soils mostly exists in biofilms. It was shown (Chathwiwat et al. 2016; Fang et al. 2012; Rodrigues and Elimelech 2010) that 99% of bacteria in the natural environment are localized in biofilm matrix. It is further known that biofilms are complex communities of one or several types of interconnected microorganisms, which are held together by an extracellular polymeric substance (EPS) (Basnet et al. 2016 ). Along with the fact that bacterial communities are considered to have a pathogenicity factor, recent studies have shown that they play other roles as well. Biofilms can be involved in various catalytic processes as living catalyst systems (Yin et al. 2019); soil biofilms play an important role in biogeochemical processes in soils (Kocur et al. 2016; Jamei, Khosravi-Nikou and Anvaripour 2013). A small number of studies are focused on the effect of silver, gold, palladium, and iron NPs on the formation of biofilms. The number of studies on the effect of ZVI NPs on the formation of biofilms is even smaller. Since the composition of a soil biofilm can include various types of bacteria, protozoa, fungi, and algae, which effectively interact with each other, the following control microorganisms were selected for our studies: *Escherichia coli* (ATCC 25922) representing Gram negative bacteria, *Pseudomonas aeruginosa* (ATCC 15692), *Enterococcus fecalis* (ATCC 29122), Proteus vulgaris (laboratory isolates), *Staphylococcus aureus* (ATCC 29213) representing Gram positive bacteria, *Bacillus cereus* (DSMZ 4312), *Bacillus subtilis* (ATCC 6633) used as a soil microorganism, and *Candida albicans* (ATCC 10231) selected as a representative of fungal microorganisms. Our research is focused on studying the effect of ZVI NPs of different morphologies on biofilm formation during their growth.

## 2. Materials and Methods

### 2.1. Chemicals

Laboratory grade high purity chemicals, such as Iron(II) sulfate, $FeSO_4$ (purity $\geq$ 99.7%), saturated iron(III) chloride, $FeCl_3$ solution (purity $\geq$ 99%), poly(N-vinylpyrrolidone) $(C_6H_9NO)_n$ (PVP, MW: 40,000 avg.), and sodium borohydride, $NABH_4$ were purchased from Sigma-Aldrich, USA. Methanol (purity $\geq$ 99.9 %) and acetone (purity $\geq$99.9 %) used in conjunction with synthesis processes were also purchased from Sigma-Aldrich, USA. De-ionized water ($>\sim$18 M$\Omega$) was used for the entire experimentation. All chemicals were used, as received and without any additional purification.

### 2.2. Nanoparticle Synthesis

The method to prepare zero-valent iron (ZVI) nanoparticles using $FeSO_4$ and $FeCl_3$, is reported by Han et al. 2016, but was modified to include a chemical reduction procedure. Low-molecular-mass PVP was used as the stabilizer. Synthesis was conducted at 15˚C in an argon atmosphere under constant stirring for 4 hrs. Nanoparticles were synthesized with an ultrasonic treatment ($Fe^0$/PVP-US NPs (A1)) and without using ultrasonic treatment ($Fe^0$/PVP NPs (A2)). The black powder of nZVI, prepared as described above, was separated from the host solution, washed with acetone and ethanol, and dried at 100°C.

### 2.3. Characterization

The zero-valent iron (ZVI) nanoparticles, synthesized as described above were studied by Fourier-transform infrared (FTIR) spectroscopy using a Perkin Elmer spectrometer model Spectrum 100 - an optical system with data collection and having spectral range of 370-7800 cm$^{-1}$ using KBr pellets in sample compartment. A PANalytical Empyrean diffractometer, using CuK$_\alpha$ radiation ($\lambda$ = 1.936 Å) was used at an accelerating voltage of 45 kV and ~40 mA current for X-ray diffraction analysis in a range of 2$\theta$ = 10°-80° (with 2$\theta$ linearity better than ±0.01), at room temperature. SEM images were observed using a Quanta 200 electronic microscope operating at 30 kV with secondary and backscattering electrons in a high vacuum mode, and TEM studies were performed using a JEOL JEM-2100F instrument. The ultrasonic treatment unit used for processing ZVI-NP had the following parameter specification: an ultrasonic bath (ISOLAB Laborgeräte GmbH), an ultrasonic power of 120 W, a heating power of 180 W, and a frequency of 40 kHz. Colloidal solutions of $Fe^0$/PVP-S1US NPs (sample 1) and $Fe^0$/PVP NPs (sample 2) in concentrations of 20 mg/mL were dissolved in dimethylsulfoxide (DMSO). The resulting colloidal solutions of iron-containing nanoparticles and microorganisms were mixed and incubated in accordance with a procedure described below.

*2.4. Cell Biology - Measurement of Antimicrobial Activity*

The antimicrobial activity test was performed for nine control microorganisms: Escherichia coli (ATCC 25922) representing Gram negative bacteria, *Pseudomonas aeruginosa* (ATCC 15692), *Enterococcus fecalis* (ATCC 29122), *Proteus vulgaris* (laboratory isolates), *Staphylococcus aureus* (ATCC 29213) representing Gram positive bacteria, *Bacillus cereus* (DSMZ 4312), *Bacillus subtilis* (ATCC 6633) used as a soil microorganism, and *Candida albicans* (ATCC 10231) selected as a representative of fungal microorganisms. The tests were conducted by using the agar well method on Mueller Hinton Agar (MHA) for bacterial control strains and on Sabouraud dextrose agar (SDA) for fungal strain.

Fresh cultures of strains were set to 0.5 McFarland [about 108 cfu (colony forming unit)/mL for *E.coli*] by using a BD CrystalSpec$^{TM}$ nephelometer (Bacton Dickinson, United States) in 0.85% (w/v) sterile physiologic serum (SPS) for the initial cell suspension of control microorganism strains. One hundred microliters of each bacterial suspension were inoculated on MHA with a sterile swab. $Fe^0$, $Fe^0$/PVP-US (sample A1), and $Fe^0$/PVP NPs (sample A2) synthesized by three different methods (A1, A2) were dissolved in dimethylsulfoxide (DMSO) to 20 mg/mL as a stock solution. After dissolving in stock solution, 100 µL of the stock solution were poured into wells on Petri dishes. The inoculated Petri dishes were incubated at 37°C for bacterial strains and 30°C for fungal strain for 24 h. After the incubation period, the inhibition zones around the agar wells (if any) were measured and recorded in millimeters (mm). In all experiments, Chloramphenicol (C30) and nystatin were used as a positive control, while DMSO was used as a negative control.

*2.5. Biofilm Formation Assay*

All of the control bacterial strains were grown in 5 mL of a LB medium at 30°C overnight. Biofilm production assays were performed using a tryptic soy broth (TSB) medium with 1% glucose. After a 24-h incubation, fresh broth cultures of each bacteria were adjusted to 0.5 McFarland in 0.85% (w/v) sterile physiologic serum using a nephelometer. Stock solutions of all substances were prepared in DMSO to a concentration of 20 mg/mL. Fresh microorganism cultures (0.5 McFarland) were directly mixed with 100 µL of $Fe^0$ NPs (A1, A2 10 mg/mL) and TSB (1% glucose) in 96 well plates, which were incubated for 72 hrs. at 37°C. Bacteria cultures in a TSB medium with 1% glucose and in a DMSO/blank medium were used as positive and negative controls, respectively. All plates were washed with tap water and air dried after 72 hrs. All wells were filled with 30% acetic acid solutions (200 µL) and incubated at room temperature. Finally, the resulting biofilms were measured by using a microtiter plate reader at an optical density of 600 nm (OD600). Results were estimated with respect to control samples.

### 3. Results and Discussion

Using XRD, SEM, and TEM, it has been shown that the synthesis with an ultrasonic treatment leads to the formation of iron NPs with a magnetic core size of 12.4Å; in this case, the iron oxide layer and the polymer layer have a thickness of 1.5–2 and 18–20 nm, respectively; the NPs are distributed discretely.

**Figure 1.** *TEM image of the morphology of (a): Fe0/PVP-US and (b): Fe0/PVP nanoparticles.*

The NPs synthesized with the use of an ultrasonic treatment are discrete spherical agglomerates distributed in the polymer matrix. Without ultrasonic treatment, iron NPs with a core size of 1.24 Å are agglomerated into aggregates of 20–200 nm; in this case, the iron oxide layer and the polymer layer have a thickness of 2–4 and 10–12 nm, respectively. Figure 2 shows the morphology of the Fe°/PVP-US NP sample synthesized in the presence of PVP with clearly visible globules composed of NPs surrounded by PVP.

**Figure 2.** *SEM image of the morphology of Fe0/PVP-US nanoparticles. PVP stabilizing ZVI NPs forms layers with a thickness of 3–5 nm for (a): Fe0/PVP-US-A1 and (b): 20–25 nm for Fe0/PVP-A2.*

The nanoparticles synthesized without ultrasonic treatment are characterized by the formation of ferromagnetic chain structures with a fragment size of 200 nm, which are shown in Figure 3, as a skeleton architecture consisting of chips.

**Figure 3.** *SEM image of the morphology of Fe0/PVP NPs under different magnifications and recording parameters: (a): Mag = 10.00K X, 1 μm, (b), Mag = 140.16K X 100 nm*

Figure 4 shows diffraction patterns of the nanopowder of $Fe^0$/PVP-US and $Fe^0$/PVP nanoparticles. The main diffraction peaks correspond to the crystallographic planes of cubic inverted spinel (space group *Fd3m*, a = 8.3952 Å). Size $d$ of the $Fe^0$ crystallites is $d = 12 \pm 1$ Å. The results are in good agreement with the spectra from the diffractograms database (Costerton et al. 1987), thereby confirming crystallinity of the synthesized iron nanopowder and nanoparticles. Size $d$ of the $Fe^0$ crystallites was calculated using the half-width of the diffraction peaks by the Scherer formula (Patterson, 1939):

$$d = k \cdot \lambda / (\beta \cos \theta), \quad (1)$$

where k (=1) is a dimensionless shape factor, $\lambda = 1.93604$ Å is the X-ray wavelength, $\beta$ is the line broadening at half the maximum intensity (FWHM) and $\theta$ is the Bragg diffraction angle. According to eq. 1, the sizes of these crystallites of $Fe^0$/PVP-US and $Fe^0$/PVP nanoparticles were calculated to be 12.0 and 12.4 Å, respectively, as per X-ray diffraction line width.

**Figure 4**: *XRD pattern of (a) -Fe0/PVP-US and (b) - Fe0/PVP nanoparticles.*

*Figure 5: FTIR spectrum of (a) the Fe0/PVP-US and (b) Fe0/PVP NPs*

Results of FTIR spectroscopic studies of the samples (FTIR spectra of the $Fe^0$/PVP-US and $Fe^0$/PVP NPs are shown in (Fig. 5). The absorption band around 2054-2302 $cm^{-1}$ is attributed to the presence of $CO_2$ molecules in the air (Kafayati et al. 2013). FTIR spectroscopy was used to investigate the interaction of PVP with iron NPs. Analysis of the FTIR spectra of the $Fe^0$/PVP-US and $Fe^0$/PVP samples, as shown in Figure 5, suggests that the polymers under investigation are coordinated for dimensionality using chelation by binding via oxygen atoms of the PVP ring and iron atoms, and also by binding via nitrogen atoms, as appears to be the case for gold nanoparticles, according to Kiran et al. (2014). The shift of the absorption band of the C=O bond from 1694 $cm^{-1}$ for the pure polymer (Figure 5) to 1735 $cm^{-1}$ is usually representative of the bond and the formation of a C=O–Me (metal) bond (Wu et al. 2019). In this case, a larger shift to the blue region is observed for the samples synthesized under an ultrasonic treatment; which is associated with a more highly ordered packing and is consistent with the microscopy data, as shown in Figure 2. At 500-600 $cm^{-1}$, the observed peak results from the Fe-O bond, which for this investigation, is evident by a small peak in the region of 580 $cm^{-1}$ and is attributed to $Fe_2O_3$ present in (a): $Fe^0$/PVP-US NPs, and in the region of 564 $cm^{-1}$ to (b): $Fe^0$/PVP NPs, formed during the synthesis.

*Figure 6: Sensitivity of control strains to Fe0/PVP NPs synthesized without an ultrasonic treatment – (a): positive control (Nistatin) (b): positive control (Chloramphenicol)*

During synthesis, the PVP was used as a polymeric stabilizer for the fabrication of iron NPs. The introduction of polymer during synthesis, facilitates formation of a negatively charged carbonyl group on the surface of the iron NPs, which contribute to the formation of a stable colloidal solution. The presence of large PVP molecules creates manifestation of a steric effect that prevents their aggregation.

Colloidal solutions based on the synthesized NPs were used in several biotests in this investigation. According to the antimicrobial tests, the results of experiments conducted on two types of NPs showed no antibacterial effects. Furthermore, effects of $Fe^0$ NPs have been studied on the biofilm formation ability of ATCC control strains. According to the assays, three $Fe^0$ NPs synthesized by several different methods do not show any negative effect on the formation of biofilms of microbial strains compared with their positive controls, as shown in Figure 6 and Figure 7. In Figure 6, sensitivity of control strains to $Fe^0$/PVP NPs synthesized without an ultrasonic treatment is shown with positive control using Nistatin Chloramphenicol, as compared to Figure 7, shown with ultrasonic treatment.

The highest absorbance values of the tested micro-organisms in control wells are found for *S. aureus, B. cereus, P. aeruginosa*, and *C. albicans*. Other test microorganisms, such as *E.coli, B. subtilis, E. faecalis, K. pneumoniae*, and *P. vulgaris*, have lower biofilm absorbance values. In addition, the estimation of the results concerning specific microorganisms has shown that A2 is the most effective in the formation of biofilms of *E.coli, K pneumoniae*, and *C. albicans* of all the tested NPs. The *C. albicans* formed biofilm, is the best characteristics. The $Fe^0$/PVP-A2 NPs have an inhibitory effect on all the test microorganisms (Figure 8). Considering the fact that biofilm formation is a preferred architecture in nature, in this study, the synthesized $Fe^0$/PVP-US-A1 NPs are not antibacterial and contribute to biofilm formation.

*Figure 7:* *Sensitivity of control strains to Fe0/PVP-US NPs synthesized under an ultrasonic treatment. (a): Positive control (Nistatin), (b): positive control (Chloramphenicol)*

Apparently, different results of the bioassay on biofilm formation can primarily be attributed to the difference in the ZVI NP synthesis methods: with and without an ultrasonic treatment of the synthetic solution; the different methods lead to the formation of NPs with different morphologies, which are shown in micrographs in Figures. 2 and 3, and sensitivity control images in Figure 6 and 7.

In tests with $Fe^0$/PVP-US-A1 NPs having a more highly ordered structure, a core of 12.0 Å, smaller aggregates of 10–15 nm, and a higher surface area, which is characteristic of nanoparticles prepared using an ultrasonic treatment, the growth of the biofilm is inhibited to a greater extent. According to Koczkur et al. 2015; Postolachi et al. 2019; and Sidorenko et al. 2020, ZVI NPs exhibit a high reactivity; accordingly, ZVI NPs can be involved in the Fenton reaction [19] to form of active oxygen species. Active oxygen species have a detrimental effect on cell membranes; this fact was noted in our earlier studies for soil microorganisms (Fang et al. 2012) and in studies of Simonin and Richaume (2015). As a consequence, A1 NPs have a smaller effect on the formation of biofilms of the tested microorganisms, as shown in Figures 8a and 8b.

It is known that the formation of biofilms occurs in several stages; one of them is adhesion, i.e., the fixation of microorganisms; it is obvious that the developed surface of skeleton magnetic aggregates of NPs prepared without exposure to ultrasound is preferable taking into account the properties (such as charge state and dimension) of each of the microorganisms. Thus, given the fact that biofilm formation is the preferred architecture in nature, the A2 NPs synthesized in this study are not antibacterial; they contribute to the formation of biofilms.

*Figure 8:* *Effect of Fe0/PVP-US-A1 (a) and Fe0/PVP-A2 Nps (b) synthesized by different methods on the formation of a biofilm of control strains.*

It is also very important to use a stabilizer forming ZVI NPs. The stabilizer generally is water-soluble polymer PVP and is commonly used in the synthesis of NPs owing to the unique properties. This offers an interesting possibility of preparing stable aqueous colloids, as evidenced by several studies by Ikuma, Decho and Lau (2015) and Vaseashta (2015). (N-vinylpyrrolidone) stabilizing ZVI NPs exhibits high adhesion properties; therefore, it can contribute to an increase in the aggregation of microorganisms into a biofilm. In addition, a larger polymer layer is formed on NPs prepared without ultrasonic treatment, although according to IR spectroscopy, the polymer layer formed with the use of an ultrasonic treatment is more coordinated.

## 4. Conclusions

In this investigation, ZVI NPs were synthesized by the reduction of iron salts and stabilized with a low-molecular-mass hydrophilic polymer (PVP) with and without exposure to an ultrasonic treatment. The morphology of ZVI NPs was determined using the SEM and TEM methods. The magnetic core size is almost identical for Fe/PVP-US Fe/PVP and Fe/PVP; it is 12.0 and 12.4 Å, respectively. The NPs synthesized with the use of an ultrasonic treatment are discrete spherical agglomerates distributed in the polymer matrix. The NPs synthesized without ultrasonic treatment are characterized by the formation of ferromagnetic chain structures with a fragment size of 200 nm. FTIR spectroscopy has revealed the formation of a C–O–Fe bond for iron NPs synthesized using an ultrasonic treatment.

Colloidal solutions based on the synthesized $Fe^0$/PVP-US and $Fe^0$/PVP NPs have been used to study the formation of biofilms during their growth and determine the antimicrobial activity. It has been shown that the formation of biofilms of various microorganisms is obviously affected by the methods used to synthesize the NPs and, accordingly, the morphology of the resulting NPs, dimension, and coordination of the stabilizer. The results have shown that $Fe^0$ NPs, which have a stimulating effect on biofilm formation in control microorganisms, should be studied with other agriculture-friendly Plant Growth-Promoting Rhizobacteria (PGPRs) to determine the possibilities of their use in agriculture for future prospects, especially due to their bioremediation nature. Hence, it is concluded that the use of ZVI NPs for bioremediation is an emerging and essential field, playing an increasingly important role in addressing innovative and effective solutions to a vast range of environmental challenges.

**Acknowledgments:** The work is done in framework of the project #65/22.10.19A, titled "Nanostructuri și nanomateriale funcționale pentru industrie și agricultură", with partial support from the SPINTECH project, under grant agreement #810144.

**Funding:** This research was funded with partial support of the SPINTECH project under joint grant agreement # 810144.

**Conflicts of Interest:** The authors declare no conflict of interest.